# Quantum State Reduction: An Operational Approach

Masanao Ozawa

*School of Informatics and Sciences, Nagoya University, Nagoya 464-01, Japan*

**Abstract**

A rigorous theory of quantum state reduction, the state change of the measured system caused by a measurement conditional upon the outcome of measurement, is developed fully within quantum mechanics without leading to the vicious circle relative to the von Neumann chain. For the basis of the theory, the local measurement theorem provides the joint probability distribution for the outcomes of local successive measurements on a noninteracting entangled system without assuming the projection postulate, and the quantum Bayes principle enables us to determine operationally the quantum state from a given information on the outcome of measurement.

1. Introduction

A measuring process consists of two stages: the first stage is the interaction between the object and the apparatus, which transduces the measured observable to the probe observable, and the second stage is the detection of the probe observable, which amplifies the probe observable to a directly sensible macroscopic variable without further disturbing the object. Any measurement on a microscopic system changes the state of the measured system. There are two ways of describing the state change caused by a measurement. One is *selective state change*, or so-called *state reduction*, which depends probabilistically on the outcome of measurement. The other is *nonselective state change*, which does not depend on the outcome of measurement. The conceptual status of state reduction is still polemical in measurement theory [1, 2, 3], whereas the nonselective state change is well described by the open-system dynamics in the first stage. One of major problems concerning the notion of state reduction is whether the second stage plays any role in changing the state of the object dynamically.

Accepting that every measurement accompanies the interaction between the object and the apparatus, one can expect that the rule of state reduction can be derived from the Schrödinger equation for the composite system of the object and the apparatus. According to the orthodox view, the Schrödinger equation for the composite system merely transforms, however, the problem of a measurement on the object to the problem of an observation on the apparatus, but in order to derive the state reduction in the object one still needs the rule of state reduction, or the projection postulate, for the composite system [4, p. 329]. Thus, the program of deriving the rule of state reduction from the Schrödinger equation for the composite system has been considered to fall into a vicious circle sometimes called von Neumann's chain [5, Section 11.2].

Consider a measurement of an observable $A$ of a quantum system $\mathbf{S}$ described by a Hilbert space $\mathcal{H}_{\mathbf{S}}$. Suppose that at the time of measurement the object is in the state (density operator) $\rho$, the apparatus is in the state $\sigma$, and the time evolution of the object-apparatus composite system during the interaction is represented by a unitary operator $U$ on the tensor product Hilbert space $\mathcal{H}_\mathbf{S} \otimes \mathcal{H}_\mathbf{A}$, where $\mathcal{H}_\mathbf{A}$ is the Hilbert space of the apparatus. Then the state $\rho'$ of the object just after measurement is obtained by the partial trace $\mathrm{Tr}_\mathbf{A}$ over the Hilbert space $\mathcal{H}_\mathbf{A}$ of the apparatus as follows:

$$\rho' = \mathrm{Tr}_\mathbf{A}[U(\rho \otimes \sigma)U^\dagger]. \tag{1}$$

This formula determines the nonselective state change $\rho \mapsto \rho'$. On the other hand, the state reduction $\rho \mapsto \rho_a$ conditional upon the outcome $a$ is related to the nonselective state change by

$$\rho' = \sum_a P(a)\rho_a, \tag{2}$$

where $P(a)$ is the probability of obtaining the outcome $a$.

The conventional derivation of the state reduction from a given model of measuring process is to compute the state of the object-apparatus composite system just after the first stage assuming the Schrödinger equation for the composite system and to apply the projection postulate to the subsequent probe detection [6, 7, 8, 9, 10]. Thus, the state $\rho_a$ is given by

$$\rho_a = \frac{\mathrm{Tr}_\mathbf{A}[(1 \otimes E^B(a))U(\rho \otimes \sigma)U^\dagger(1 \otimes E^B(a))]}{\mathrm{Tr}[(1 \otimes E^B(a))U(\rho \otimes \sigma)U^\dagger(1 \otimes E^B(a))]} \tag{3}$$

where $B$ is the probe observable and $E^B(a)$ is the projection operator with the range $\{\psi \in \mathcal{H}_\mathbf{A} |\ B\psi = a\psi\}$.

The application of the projection postulate in the above derivation represents the state change in the second stage, which includes another *interaction* between the probe (i.e., the microscopic subsystem of the apparatus having the probe observable) and the macroscopic part of the apparatus measuring the probe observable. Thus, the conventional derivation assigns the second stage part of the dynamical cause of the state reduction.

The validity of the above derivation is, however, limited or even questionable, besides the interpretational questions of the vicious circle mentioned above, because of the following reasons:

1. The probe detection, such as photon counting, in some measuring apparatus does not satisfy the projection postulate [11].

2. When the probe observable has continuous spectrum, the projection postulate to be applied cannot be formulated properly in the standard formulation of quantum mechanics [12].

3. When another measurement on the same object follows immediately after the first measurement, the measuring apparatus for the second measurement can interact with the object just after the first stage and just before the second stage of the first measurement. The state reduction obtained by the conventional approach, which determines the state just after the second stage of the first measurement, cannot give the joint probability distribution of the outcomes of the above consecutive measurements [13].

Despite the above points, it is usually claimed that state reduction has not yet occurred at the first stage but needs the further interaction between the apparatus and the observer's ego [14], between the apparatus and the environment [15, 16, 17], or between the probe and the macroscopic detector [18, 19]—the application of the projection postulate to the object-apparatus composite system is considered to be an *ad hoc* expression of this kind of interaction.

In this paper, I will present a consistent argument which derives the state reduction from the Schrödinger equation for the object-apparatus composite system without appealing to the projection postulate for the probe measurement. For the basis of the theory, the local measurement theorem provides the joint probability distribution for the outcomes of local successive measurements on a noninteracting entangled system without assuming the projection postulate, and the quantum Bayes principle enables us to determine operationally the quantum state from a given information on the outcome of measurement. They will naturally lead to the state reduction for an arbitrary model of measurement.

For simplicity we will be confined to measurements of *discrete observables*, but it will be easy for the reader to generalize the argument to continuous observables and to join the argument to the general theory developed in such papers as [12, 20, 21, 22, 23, 24, 25, 26].

## 2. States and probability distributions

Let **S** be a quantum system described by a Hilbert space $\mathcal{H}$. The statistical formula for an observable $A$ of **S** is stated as follows: *If the system **S** is in the state $\rho(t)$ at time $t$, then the probability, $\Pr\{A(t) = a\}$, that the measurement of an observable $A$ at the time $t$ leads to the outcome a is given by*

$$\Pr\{A(t) = a\} = \mathrm{Tr}[E^A(a)\rho(t)]. \tag{4}$$

Let $H$ be the Hamiltonian of **S**. The solution of the Schrödinger equation is stated as follows: *If the system **S** is in the state $\rho(t)$ at time $t$ and if the system **S** is isolated between the time $t$ and $t + \tau$, then at the time $t + \tau$ the system is in the state*

$$\rho(t+\tau) = e^{-iH\tau/\hbar}\rho(t)e^{iH\tau/\hbar}. \tag{5}$$

When the state in the past is known, these formulas are used for predicting the probability distribution of the outcome of a future measurement. It is possible, however, to determine conversely the state of the system when the probability distributions are given in advance. Thus, we are able to reformulate these formulas into the following one principle about the relation between the notion of state and the set of the probability distributions of the outcomes of the measurements of observables.

**Rule 1.** *The system **S** is in the state $\rho(t)$ at time $t$ if and only if the following statement holds: if the system **S** is isolated from the time $t$ to $t + \tau$ with $0 \leq \tau$, and if an arbitrary observable $X$ of **S** is measured at the time $t + \tau$, then the probability distribution $\Pr\{X(t+\tau) = x\}$ of the outcome of this measurement is given by*

$$\Pr\{X(t+\tau) = x\} = \mathrm{Tr}[E^X(x)e^{-iH\tau/\hbar}\rho(t)e^{iH\tau/\hbar}]. \tag{6}$$

**Remark.** In the above rule, we assume that the validity of the statement that the system **S** is in the state $\rho(t)$ at time $t$ is independent of whether the observable $X$ is actually measured at the future time $t + \tau$. Moreover, we do not assume that any system is in some state at any time.

It can be seen easily that Rule 1 can be derived from the Schrödinger equation and the statistical formula and that the rule brings logically no new element to quantum mechanics. In fact, if the system **S** is in the state $\rho(t)$, then (6) is a consequence of (4) and (5). Since (6) is satisfied by at most one density operator, the rule determines the state $\rho(t)$ uniquely.

## 3. Composite systems

Another two basic principles of quantum mechanics is necessary when we treat more than one system. The rule for representing composite systems is stated as follows.

**Rule 2.** *Let $\mathbf{S}_{12}$ be the composite system of two subsystems $\mathbf{S}_1$ and $\mathbf{S}_2$. If the systems $\mathbf{S}_1$ and $\mathbf{S}_2$ are described by the Hilbert spaces $\mathcal{H}_1$ and $\mathcal{H}_2$ respectively, the composite system $\mathbf{S}_{12}$ is described by their tensor product $\mathcal{H}_1 \otimes \mathcal{H}_2$, the observable of $\mathbf{S}_1$ represented by the operator $X$ on $\mathcal{H}_1$ is represented by $X \otimes 1$ on $\mathcal{H}_1 \otimes \mathcal{H}_2$, and the observable of $\mathbf{S}_2$ represented by the operator $Y$ on $\mathcal{H}_2$ is represented by $1 \otimes Y$ on $\mathcal{H}_1 \otimes \mathcal{H}_2$.*

The following principle is plausible from our belief in local universal systems.

**Rule 3.** *For any system $\mathbf{S}_1$ and any time interval from $t$ to $t + \tau$, there is an external system $\mathbf{S}_2$ in a state $\rho_{12}(t)$ at the time $t$ such that $\mathbf{S}_1 + \mathbf{S}_2$ is isolated from the time $t$ to $t + \tau$. (The system $\mathbf{S}_2$ may depend on the time interval and may be an empty system.)*

**Theorem 1.** *If the composite system $\mathbf{S}_1 + \mathbf{S}_2$ is in the state $\rho_{12}(t)$ at time $t$, then the state $\rho_1(t)$ of the system $\mathbf{S}_1$ at the time $t$ exists and is given by*

$$\rho_1(t) = \mathrm{Tr}_2[\rho_{12}(t)], \tag{7}$$

*where $\mathrm{Tr}_2$ is the partial trace operation over the Hilbert space $\mathcal{H}_2$.*

*Proof.* Let $\mathbf{S}_3$ be an arbitrary system external with $\mathbf{S}_{12}$ such that the composite system $\mathbf{S}_{123} = \mathbf{S}_{12} + \mathbf{S}_3$ is in the state $\rho_{123}(t)$ at the time $t$. Suppose first that an arbitrary observable $X$ of $\mathbf{S}_1$ is measured at the time $t$. Then by Rule 1 and Rule 2, we have

$$\Pr\{X(t) = x\} = \mathrm{Tr}[(E^X(x) \otimes 1)\rho_{12}(t)] = \mathrm{Tr}[(E^X(x) \otimes 1 \otimes 1)\rho_{123}(t)].$$

Since $X$ is arbitrary, we have

$$\mathrm{Tr}_{23}[\rho_{123}(t)] = \mathrm{Tr}_2[\rho_{12}(t)], \tag{8}$$

where $\text{Tr}_{23}$ stands for the partial trace over the Hilbert space $\mathcal{H}_2 \otimes \mathcal{H}_3$. Next, suppose that the system $\mathbf{S}_1$ is isolated from time $t$ to $t+\tau$ with $0 \leq \tau$ and that an arbitrary observable $X$ of $\mathbf{S}_1$ is measured at the time $t+\tau$. Let $\mathbf{S}_3$ be a system external with $\mathbf{S}_{12}$ such that the composite system $\mathbf{S}_{123} = \mathbf{S}_{12} + \mathbf{S}_3$ is in the state $\rho_{123}(t)$ at the time $t$ and is isolated from the time $t$ to $t+\tau$. There is at least one such system by Rule 3. Let $U_{123}(t, t+\tau)$ be the time evolution operator of the system $\mathbf{S}_{123}$ from $t$ to $t+\tau$. Then by Rule 1 and Rule 2, we have

$$\Pr\{X(t+\tau) = x\} = \text{Tr}[(E^X(x) \otimes 1 \otimes 1) U_{123}(t, t+\tau) \rho_{123}(t) U_{123}(t, t+\tau)^\dagger]. \quad (9)$$

Since $\mathbf{S}_1$ is isolated from the time $t$ to $t+\tau$, there is no interaction between $\mathbf{S}_1$ and $\mathbf{S}_2 + \mathbf{S}_3$, so that we have

$$U_{123}(t, t+\tau) = e^{-iH_1 \tau/\hbar} \otimes U_{12}(t, t+\tau), \quad (10)$$

where $H_1$ is the Hamiltonian of $\mathbf{S}_1$ and $U_{12}(t, t+\tau)$ is the time evolution operator of $\mathbf{S}_2 + \mathbf{S}_3$ from $t$ to $t+\tau$. By (9) and (10), we have

$$\Pr\{X(t+\tau) = x\} = \text{Tr}[E^X(x) e^{-iH_1 \tau/\hbar} \text{Tr}_{23}[\rho_{123}(t)] e^{iH_1 \tau/\hbar}],$$

and by (8) we have

$$\Pr\{X(t+\tau) = x\} = \text{Tr}[E^X(x) e^{-iH_1 \tau/\hbar} \text{Tr}_2[\rho_{12}(t)] e^{iH_1 \tau/\hbar}]. \quad (11)$$

Since $X$ and $\tau$ are arbitrary, Rule 1 concludes from (11) that the system $\mathbf{S}_1$ is in the state $\rho_1(t) = \text{Tr}_2[\rho_{12}(t)]$ at the time $t$. □

**Remark.** According to the above proof, (11) holds even when the system $\mathbf{S}_2$ interacts with other systems before the time $t + \tau$.

## 4. Local Measurement Theorem

Let $\mathbf{S}_1$ be a quantum system described by a Hilbert space $\mathcal{H}_1$ with the Hamiltonian $H_1$ and let $\mathbf{S}_2$ be a quantum system described by a Hilbert space $\mathcal{H}_2$ with the Hamiltonian $H_2$. Let $A$ be an observable of the system $\mathbf{S}_1$ and $B$ an observable of the system $\mathbf{S}_2$. Suppose that the composite system $\mathbf{S}_1 + \mathbf{S}_2$ is originally in the state $\rho_{12}$. Suppose that the observable $A$ is measured at time $t_1$, the observable $B$ is measured at time $t_2$ with $0 < t_1 < t_2$, and that there is no interaction between $\mathbf{S}_1$ and $\mathbf{S}_2$.

If the $A$-measurement satisfies the projection postulate, the joint probability distribution of the outcomes of the $A$-measurement and the $B$-measurement is given by

$$\begin{aligned}\Pr\{A(t_1) &= a, B(t_2) = b\} \\ &= \text{Tr}[(e^{iH_1 t_1/\hbar} E^A(a) e^{-iH_1 t_1/\hbar} \otimes e^{iH_2 t_2/\hbar} E^B(b) e^{-iH_2 t_2/\hbar}) \rho_{12}]. \end{aligned} \quad (12)$$

In what follows we shall derive the above formula *without* assuming the projection postulate. Suppose that the apparatus measuring the observable $A$ is described by a quantum system $\mathbf{A}$ with the Hilbert space $\mathcal{H}_\mathbf{A}$. We say that the measurement of $A$ is *local* at the system $\mathbf{S}_1$ if the measuring interaction for this measurement occurs only between the apparatus $\mathbf{A}$ and the system $\mathbf{S}_1$, or precisely, if the operator representing the measuring interaction commutes with every observable of $\mathbf{S}_2$. If this is the case, the total Hamiltonian of the composite system $\mathbf{A} + \mathbf{S}_1 + \mathbf{S}_2$ during the measuring interaction is represented by

$$H_{tot} = H_\mathbf{A} \otimes 1 \otimes 1 + 1 \otimes H_1 \otimes 1 + 1 \otimes 1 \otimes H_2 + K H_{int} \otimes 1, \qquad (13)$$

where $H_\mathbf{A}$ is the Hamiltonian of the apparatus, $H_{int}$ is the operator on $\mathcal{H}_\mathbf{A} \otimes \mathcal{H}_1$ representing the measuring interaction, and $K$ is the coupling constant. Then, we can prove the following theorem in the standard formulation of quantum mechanics [27].

**Theorem 2. (Local Measurement Theorem)** *If the measurement of A is local at the system $\mathbf{S}_1$, the joint probability distribution of the outcomes of the A-measurement and the B-measurement is given by (12).*

## 5. Quantum Bayes Principle

Suppose that we are given the joint probability distribution of two (discrete) random variables $X, Y$. The prior distribution of $X$ is equal to the marginal distribution of $X$. If one measures $Y$, the *information* "$Y = y$" changes the probability distribution of $X$ for any outcome $y$. The posterior distribution of $X$ is defined as the conditional probability distribution of $X$ given $Y = y$. The principle of changing the probability distribution from the prior distribution to the posterior distribution is called the *Bayes principle*.

In quantum mechanics, the notion of probability distribution is related to the notion of state. As described in Rule 1, "the state of the system" is equivalent to "the probability distributions of all the observables of the system". Thus the Bayes principle yields the state change of a quantum system if the probability distributions of all the observables of the system has changed by the Bayes principle. We formulate this principle of state changes as follows:

**The Quantum Bayes Principle:** *If an information changes the probability distributions of all the observables of a quantum system according to the Bayes principle, then the information changes the state of the system according to the change of the probability distributions.*

## 6. Quantum Bayes Principle in Entangled Systems

In what follows we consider the composite system $\mathbf{S}_1 + \mathbf{S}_2$ originally in the state $\rho_{12}$ and the successive local measurements of $A$ in $\mathbf{S}_1$ at the time $t$ and $X$ in $\mathbf{S}_2$ at the time $t + \tau$ with $0 < \tau$. In this section, we assume that $A$ and $t$ are fixed but $X$ and $\tau$ are arbitrary.

By the Local Measurement Theorem, the joint probability distribution of the outcome $A(t)$ of the A-measurement at $t$ and the outcome $X(t+\tau)$ of the X-measurement

at $t+\tau$ is given by

$$\Pr\{A(t)=a, X(t+\tau)=x\}$$
$$= \text{Tr}[(e^{iH_1 t/\hbar}E^A(a)e^{-iH_1 t/\hbar} \otimes e^{iH_2(t+\tau)/\hbar}E^X(x)e^{-iH_2(t+\tau)/\hbar})\rho_{12}]. \quad (14)$$

Thus, the prior probability distribution of $X(t+\tau)$ is defined as the marginal distribution of $X(t+\tau)$, i.e.,

$$\begin{aligned}
\Pr\{X(t+\tau)=x\} &= \sum_a \Pr\{A(t)=a, X(t+\tau)=x\} \\
&= \text{Tr}[(1 \otimes e^{iH_2(t+\tau)/\hbar}E^X(b)e^{-iH_2(t+\tau)/\hbar})\rho_{12}] \\
&= \text{Tr}[E^X(x)e^{-iH_2\tau/\hbar}(e^{-iH_2 t/\hbar}\text{Tr}_1[\rho_{12}]e^{iH_2 t/\hbar})e^{iH_2\tau/\hbar}].
\end{aligned}$$

Since $X$ and $\tau$ are arbitrary, from Rule 1 the state of the system $\mathbf{S}_2$ at the time $t$ is given by

$$\rho_2(t) = e^{-iH_2 t/\hbar}\text{Tr}_1[\rho_{12}]e^{iH_2 t/\hbar}. \quad (15)$$

This shows that the $A$-measurement at the time $t$ does not interrupt the unitary evolution of the system $\mathbf{S}_2$ from the time 0. The state $\rho_2(t)$ describes the prior probability distributions of the outcomes of all observables $X$ of the system $\mathbf{S}_2$ after the time $t$, i.e.,

$$\Pr\{X(t+\tau)=x\} = \text{Tr}[E^X(x)e^{-iH_2\tau/\hbar}\rho_2(t)e^{iH_2\tau/\hbar}]$$

and is called the *prior state* of $\mathbf{S}_2$ at $t$.

If one reads out $A(t)$, the outcome of the $A$-measurement at $t$, the information "$A(t)=a$" changes the probability distribution of $X(t+\tau)$, the outcome of the $X$-measurement at $t+\tau$, for any outcome $a$ from the prior distribution to the posterior distribution according to the Bayes principle. The posterior distribution of $X(t+\tau)$ is defined as the conditional probability distribution of $X(t+\tau)$ given $A(t)=a$, i.e.,

$$\begin{aligned}
&\Pr\{X(t+\tau)=x|A(t)=a\} \\
&= \frac{\Pr\{A(t)=a, X(t+\tau)=x\}}{\sum_x \Pr\{A(t)=a, X(t+\tau)=x\}} \\
&= \frac{\text{Tr}[(e^{iH_1 t/\hbar}E^A(a)e^{-iH_1 t/\hbar} \otimes e^{iH_2(t+\tau)/\hbar}E^X(x)e^{-iH_2(t+\tau)/\hbar})\rho_{12}]}{\text{Tr}[(e^{iH_1 t/\hbar}E^A(a)e^{-iH_1 t/\hbar} \otimes 1)\rho_{12}]}.
\end{aligned}$$

Now, we have

$$\text{Tr}[(e^{iH_1 t/\hbar}E^A(a)e^{-iH_1 t/\hbar} \otimes e^{iH_2(t+\tau)/\hbar}E^X(x)e^{-iH_2(t+\tau)/\hbar})\rho_{12}]$$
$$= \text{Tr}[E^X(x)e^{-iH_2(t+\tau)/\hbar}\text{Tr}_1[(e^{iH_1 t/\hbar}E^A(a)e^{-iH_1 t/\hbar} \otimes 1)\rho_{12}]e^{iH_2(t+\tau)/\hbar}].$$

Thus, letting

$$\rho_2(t|A(t)=a) = \frac{e^{-iH_2 t/\hbar}\text{Tr}_1[(e^{iH_1 t/\hbar}E^A(a)e^{-iH_1 t/\hbar} \otimes 1)\rho_{12}]e^{iH_2 t/\hbar}}{\text{Tr}[(e^{iH_1 t/\hbar}E^A(a)e^{-iH_1 t/\hbar} \otimes 1)\rho_{12}]}, \quad (16)$$

we have

$$\Pr\{X(t+\tau) = x | A(t) = a\} = \text{Tr}[E^X(x)e^{-iH_2\tau/\hbar}\rho_2(t|A(t) = a)e^{iH_2\tau/\hbar}]. \quad (17)$$

Since $X$ and $\tau$ are arbitrary, from Rule 1 the state of the system $\mathbf{S}_2$ at the time $t$ is described by the state $\rho_2(t|A(t) = a)$, which evolves unitarily according to the system Hamiltonian of $\mathbf{S}_2$. Whereas we have attributed two different states, $\rho_2(t)$ and $\rho_2(t_2|A(t_1) = a)$, to the system $\mathbf{S}_2$ at the time $t$, they will not lead to a contradiction because they represent the two different sets of probability distributions, the prior distributions and the posterior distributions. The dualism concerning the prior and posterior probabilities is not the peculiarity of quantum mechanics but inherent in probability theory. Therefore, we should allow the dual nature of the state description derived from the dualism in probability theory and we should conclude that the information "$A(t) = a$" changes the state of the system $\mathbf{S}_2$ at the time $t$ from the *prior state* $\rho_2(t)$ to the *posterior state* $\rho_2(t|A(t) = a)$ according to the quantum Bayes principle.

## 7. Quantum state reduction based on the quantum Bayes principle

Let $\mathbf{S}$ be a quantum system described by a Hilbert space $\mathcal{H}$ with the Hamiltonian $H$, and let $\mathbf{A}$ the apparatus measuring the observable $A$ of $\mathbf{S}$. The apparatus $\mathbf{A}$ is treated as a quantum system described by a Hilbert space $\mathcal{H}_\mathbf{A}$. Suppose that the measurement of $A$ is carried out by the interaction between $\mathbf{S}$ and $\mathbf{A}$ from the time $t$, the time of measurement, to the time $t + \Delta t$, the time just after measurement, and that after $t + \Delta t$ the object is free from the apparatus. Let $\rho(t)$ be the state of $\mathbf{S}$ at the time of measurement. We assume that $\rho(t)$ is arbitrarily given. By Rule 1 the probability distribution $\Pr\{A(t) = a\}$ of the outcome of measurement is given by

$$\Pr\{A(t) = a\} = \text{Tr}[E^A(a)\rho(t)]. \quad (18)$$

Suppose that at the time of measurement, $t$, the apparatus $\mathbf{A}$ is prepared in a fixed state $\sigma$, the time evolution of the composite system $\mathbf{S} + \mathbf{A}$ from the time $t$ to $t + \Delta t$ is represented by a unitary operator $U$ on the Hilbert space of $\mathbf{S} + \mathbf{A}$, and that the outcome of measurement is obtained by measuring the probe observable $B$ in the apparatus at the time just after measurement, $t + \Delta t$. We assume that $B$ has the same spectrum as $A$ and that the outcome "$B(t + \Delta t) = a$" is interpreted as the outcome "$A(t) = a$". Thus, we have

$$\Pr\{A(t) = a\} = \Pr\{B(t + \Delta t) = a\}. \quad (19)$$

Since the system $\mathbf{S} + \mathbf{A}$ is in the state $U(\rho(t) \otimes \sigma)U^\dagger$ at the time $t + \Delta t$, from Rule 1 we have

$$\Pr\{B(t + \Delta t) = a\} = \text{Tr}[(1 \otimes E^B(a))U(\rho(t) \otimes \sigma)U^\dagger], \quad (20)$$

and hence from (18)–(20) we have

$$\text{Tr}[E^A(a)\rho(t)] = \text{Tr}[U^\dagger(1 \otimes E^B(a))U(\rho(t) \otimes \sigma)]. \quad (21)$$

Since $\rho(t)$ is arbitrary, we have

$$E^A(a) = \text{Tr}_{\mathbf{A}}[U^\dagger(1 \otimes E^B(a))U(1 \otimes \sigma)] \qquad (22)$$

for all real numbers $a$. The above relation, (22), is the sole requirement for $\sigma, U, B$ to represent an apparatus measuring the observable $A$ [12].

In what follows, we shall derive the explicit form of the state reduction of this measurement in terms of the apparatus preparation $\sigma$, the measuring interaction $U$, and the probe observable $B$ without appealing to the projection postulate for the probe measurement.

Since the composite system $\mathbf{S}+\mathbf{A}$ is $U(\rho(t)\otimes\sigma)U^\dagger$ at the time $t+\Delta t$, by Theorem 1 the state of $\mathbf{S}$ just after measurement is given by

$$\rho(t+\Delta t) = \text{Tr}_{\mathbf{A}}[U(\rho(t)\otimes\sigma)U^\dagger]. \qquad (23)$$

It follows that if an arbitrary observable $X$ of $\mathbf{S}$ is measured at the time $t+\Delta t+\tau$ with $0 \leq \tau$, by Rule 1 the probability distribution $\Pr\{X(t+\Delta t+\tau) = x\}$ of the outcome of measurement is given by

$$\Pr\{X(t+\Delta t+\tau) = a\} = \text{Tr}[E^A(a)e^{-iH\tau/\hbar}\rho(t+\Delta t)e^{iH\tau/\hbar}]. \qquad (24)$$

Since the state change $\rho(t) \mapsto \rho(t+\Delta t)$ does not depend on the outcome of the $A$-measurement at $t$, this state change is considered as the nonselective state change and the probability distribution $\Pr\{X(t+\Delta t+\tau) = x\}$ is considered as the prior distribution of the outcome $X(t+\Delta t+\tau)$ of the $X$-measurement at $t+\Delta t+\tau$.

In order to consider the state reduction, suppose that the object $\mathbf{S}$ is in the state $\rho(t+\Delta t|A(t) = a)$ at the time just after measurement provided that the measurement leads to the outcome "$A(t) = a$". If $\Pr\{A(t) = a\} = 0$, the state $\rho(t+\Delta t|A(t) = a)$ is not definite, and we let $\rho(t+\Delta t|A(t) = a)$ be an arbitrarily chosen density operator for mathematical convenience. We denote by $\Pr\{A(t) = a, X(t+\Delta t+\tau) = x\}$ the joint probability distribution of the outcome, $A(t)$, of the $A$-measurement at $t$ and the outcome, $X(t+\Delta t+\tau)$, of the $X$-measurement at $t+\Delta t+\tau$. If $\Pr\{A(t) = a\} \neq 0$, let $\Pr\{X(t+\Delta t+\tau) = x|A(t) = a\}$ be the conditional probability distribution of $X(t+\Delta t+\tau)$ given $A(t) = a$, i.e.,

$$\Pr\{X(t+\Delta t+\tau) = x|A(t) = a\} = \frac{\Pr\{A(t) = a, X(t+\Delta t) = x\}}{\Pr\{A(t) = a\}}. \qquad (25)$$

Then, under Rule 1 the state $\rho(t+\Delta t|A(t) = a)$ is naturally characterized by

$$\Pr\{X(t+\Delta t+\tau) = x|A(t) = a\} = \text{Tr}[E^X(x)e^{-iH\tau/\hbar}\rho(t+\Delta t|A(t) = a)e^{iH\tau/\hbar}] \quad (26)$$

for arbitrary $X$ and arbitrary $\tau$.

The information "A(t)=a" changes the probability distribution of $X(t+\Delta t+\tau)$ from the prior distribution $\Pr\{X(t+\Delta t+\tau) = a\}$ to the posterior distribution $\Pr\{X(t+\Delta t+\tau) = x|A(t) = a\}$ by the Bayes principle. Since the information "A(t)=a" changes the probability distribution of *all* the observables of the system $\mathbf{S}$ according to the Bayes principle, the information changes the state of $\mathbf{S}$ according to the quantum Bayes principle. Under Rule 1, the prior distribution is determined by the state $\rho(t + \Delta t)$ and the posterior distribution is determined by the state $\rho(t + \Delta t | A(t) = a)$, and hence the information "$A(t) = a$" changes the state of the object just after measurement from the prior state $\rho(t + \Delta t)$ to the posterior state $\rho(t + \Delta t | A(t) = a)$ by the quantum Bayes principle.

In order to obtain the state $\rho(t + \Delta t | A(t) = a)$ from our model of measurement specified by $(\sigma, U, B)$, suppose that at the time $t + \Delta t + \tau$ with $0 \leq \tau$ the observer were to measure an arbitrary observable $X$ of the same object **S** measured by the apparatus **A**. Naturally, the measurement of $B$ of **A** at $t + \Delta t$ is carried out *locally*, i.e., the measuring apparatus measuring $B$ of **A** does not interact with **S**. Then, by the Local Measurement Theorem we have

$$\Pr\{B(t + \Delta t) = a, X(t + \Delta t + \tau) = x\}$$
$$= \text{Tr}[(e^{iH\tau/\hbar} E^X(x) e^{-iH\tau/\hbar} \otimes E^B(a)) U(\rho(t) \otimes \sigma) U^\dagger]. \quad (27)$$

The argument in the previous section concerning the quantum Bayes principle in entangled systems can be applied to the system $\mathbf{S} + \mathbf{A}$ at the time $t + \Delta t$ as follows. By (27) the conditional probability distribution of $X(t + \Delta t + \tau)$ given $B(t + \Delta t) = a$ is obtained as

$$\Pr\{X(t + \Delta t + \tau) = x | B(t + \Delta t) = a\}$$
$$= \frac{\text{Tr}[E^X(x) e^{-iH\tau/\hbar} \text{Tr}_\mathbf{A}[(1 \otimes E^B(a)) U(\rho(t) \otimes \sigma) U^\dagger] e^{iH\tau/\hbar}]}{\text{Tr}[(1 \otimes E^B(a)) U(\rho(t) \otimes \sigma) U^\dagger]}.$$

Thus, if we let

$$\rho(t + \Delta t | B(t + \Delta t) = a) = \frac{\text{Tr}_\mathbf{A}[(1 \otimes E^B(a)) U(\rho(t) \otimes \sigma) U^\dagger]}{\text{Tr}[(1 \otimes E^B(a)) U(\rho(t) \otimes \sigma) U^\dagger]}, \quad (28)$$

we have

$$\Pr\{X(t + \Delta t + \tau) = x | B(t + \Delta t) = a\}$$
$$= \text{Tr}[E^X(x) e^{-iH\tau/\hbar} \rho(t + \Delta t | B(t + \Delta t) = a) e^{iH\tau/\hbar}]. \quad (29)$$

Since $X$ and $\tau$ are arbitrary, from Rule 1 the state of the system **S** at the time $t + \Delta t$ is described by the state $\rho(t + \Delta t | B(t + \Delta t) = a)$. Since the outcome "$B(t + \Delta t) = a$" is interpreted as the outcome "$A(t) = a$", we have

$$\Pr\{X(t + \Delta t + \tau) = x | A(t) = a\} = \Pr\{X(t + \Delta t + \tau) = x | B(t + \Delta t) = a\}. \quad (30)$$

From (29) and (30) we have

$$\Pr\{X(t + \Delta t + \tau) = x | A(t) = a\} = \text{Tr}[E^X(x) e^{-iH\tau/\hbar} \rho(t + \Delta t | B(t + \Delta t) = a) e^{iH\tau/\hbar}],$$

and hence from (26) we have

$$\rho(t + \Delta t | A(t) = a) = \rho(t + \Delta t | B(t + \Delta t) = a). \quad (31)$$

We conclude, therefore, the state reduction $\rho(t) \mapsto \rho(t + \Delta t | A(t) = a)$ is determined by

$$\rho(t + \Delta t | A(t) = a) = \frac{\text{Tr}_{\mathbf{A}}[(1 \otimes E^B(a))U(\rho(t) \otimes \sigma)U^\dagger]}{\text{Tr}[(1 \otimes E^B(a))U(\rho(t) \otimes \sigma)U^\dagger]}. \quad (32)$$

By a property of the partial trace operation, (32) is equivalent to the following equation

$$\rho(t + \Delta t | A(t) = a) = \frac{\text{Tr}_{\mathbf{A}}[(1 \otimes E^B(a))U(\rho(t) \otimes \sigma)U^\dagger(1 \otimes E^B(a))]}{\text{Tr}[(1 \otimes E^B(a))U(\rho(t) \otimes \sigma)U^\dagger(1 \otimes E^B(a))]}. \quad (33)$$

The above equation appears to be obtained from the assumption that the measurement for the composite system $\mathbf{S} + \mathbf{A}$ measuring the probe observable $B$ satisfies the projection postulate. But, such an argument does not lead to (33). Suppose that the measuring apparatus measuring $B$ interacts with the system $\mathbf{A}$ from $t + \Delta$ to $t + \Delta t + \Delta t'$. By the projection postulate, the state of the composite system $\mathbf{S} + \mathbf{A}$ at the time just after the $B$-measurement is given by

$$\rho_{\mathbf{S}+\mathbf{A}}(t+\Delta t+\Delta t'|B(t+\Delta t) = a) = \frac{(1 \otimes E^B(a))U(\rho(t) \otimes \sigma)U^\dagger(1 \otimes E^B(a))}{\text{Tr}[(1 \otimes E^B(a))U(\rho(t) \otimes \sigma)U^\dagger(1 \otimes E^B(a))]}, \quad (34)$$

provided that the $B$-measurement leads to the outcome $a$. Since (34) represents the state of the composite system at $t + \Delta t + \Delta t'$, the time just after the $B$-measurement, the partial trace of (34) represents the state of the object $\mathbf{S}$ at the time $t + \Delta t + \Delta t'$ but not at the time $t + \Delta t$. Thus, (33) is *not* a consequence of the assumption that the measurement for the composite system $\mathbf{S} + \mathbf{A}$ measuring the probe observable $B$ satisfies the projection postulate. Moreover, according to our argument (33), as well as the equivalent formula (32), holds for any way of measuring the probe observable $B$ whether it satisfies the projection postulate or not as long as it is local.